\documentclass[notitlepage,twocolumn,letterpaper,natbib,aps,prd,amsmath,amsfonts,nofootinbib,preprintnumbers,superscriptaddress,secnumarabic,groupedaddress]{revtex4-1}
\pdfoutput=1
\usepackage{amssymb,amsmath,latexsym,mathrsfs}
\usepackage{url}
\usepackage{enumitem}
\usepackage{graphicx}
\usepackage{booktabs}
\usepackage[usenames,dvipsnames]{color}
\usepackage[breaklinks,colorlinks,urlcolor=blue,citecolor=blue,linkcolor=magenta]{hyperref}
\usepackage{multirow}
\usepackage{float}
\usepackage{cases}
\usepackage{blindtext}
\usepackage{pifont}
\usepackage{hhline}
\usepackage{slashed}
\usepackage{array}
\usepackage{orcidlink}

\usepackage{xcolor}
\definecolor{linkcolor}{rgb}{0.0, 0.47, 0.75}
\definecolor{citecolor}{rgb}{1.0, 0.5, 0.0}
\hypersetup{
  linkcolor  = linkcolor,
  citecolor  = linkcolor,
  urlcolor   = linkcolor,
  colorlinks = true
}

\begin{document}

\title{The Simplest Dark Matter Model at the Edge of Perturbativity}

\author{Miguel Escudero Abenza \orcidlink{0000-0002-4487-8742}}
\email{miguel.escudero@cern.ch}
\affiliation{Theoretical Physics Department, CERN, 1211 Geneva 23, Switzerland}

\author{Thomas Hambye \orcidlink{0000-0003-4381-3119}}
\email{thomas.hambye@ulb.be}
\affiliation{Service de Physique Th\'eorique, Universit\'e Libre de Bruxelles, Boulevard du Triomphe, CP225, 1050 Brussels, Belgium}

\date{\today}

\preprint{CERN-TH-2025-087, ULB-TH/25-04}

\begin{abstract}
\noindent Increasingly sensitive direct detection dark matter experiments are testing important regions of parameter space for WIMP dark matter and pushing many models to the multi-TeV regime. This brings into question the perturbativity of these models. In this context, and in light of the new limits from the LZ experiment, we investigate the status of the simplest thermal dark matter model: a singlet scalar, real or complex, coupled to the Higgs boson. We calculate the next-to-leading order (NLO) corrections to the direct detection rates as well as for the annihilations driving thermal freeze-out. For the complex case, we find that the entire perturbative region is excluded by direct detection. For the real case we find that the mass should be $\gtrsim 20\,{\rm TeV}$ at NLO, compared with the $\gtrsim 30\,{\rm TeV}$ LO limit. We highlight that a three-fold improvement on WIMP spin independent interactions can fully test the real scalar model in the perturbative regime. Finally, for both models, there is still an allowed (albeit narrow) region near the Higgs resonance where couplings are perturbative.
\end{abstract}

\maketitle

\begin{figure*}[t]
\begin{center}
\begin{tabular}{cc}
 \includegraphics[width=0.49\textwidth]{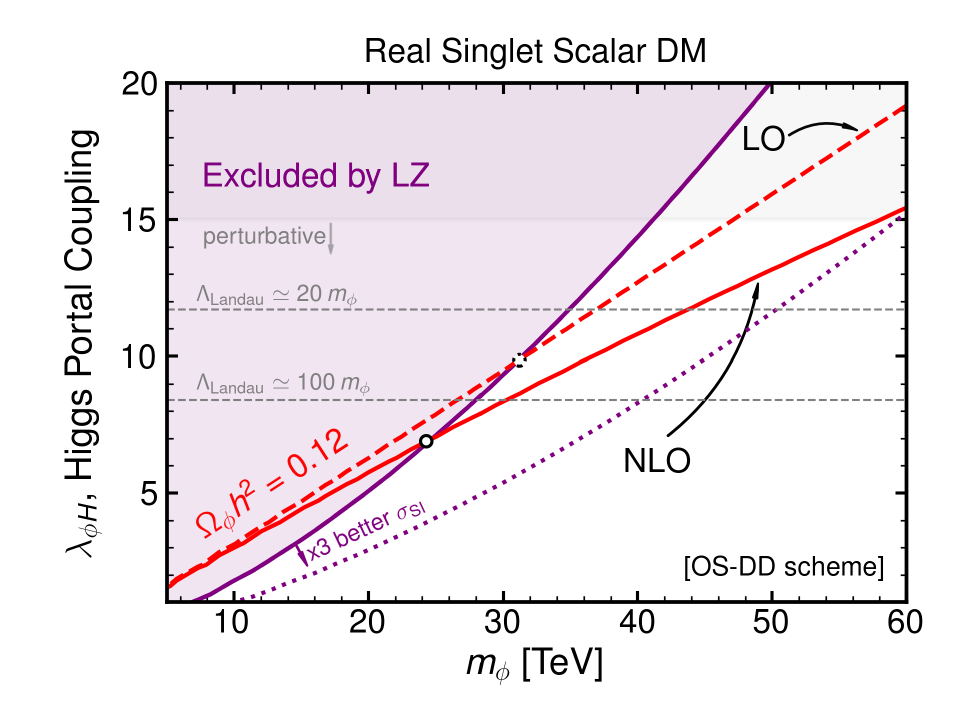}
    &   \includegraphics[width=0.49\textwidth]{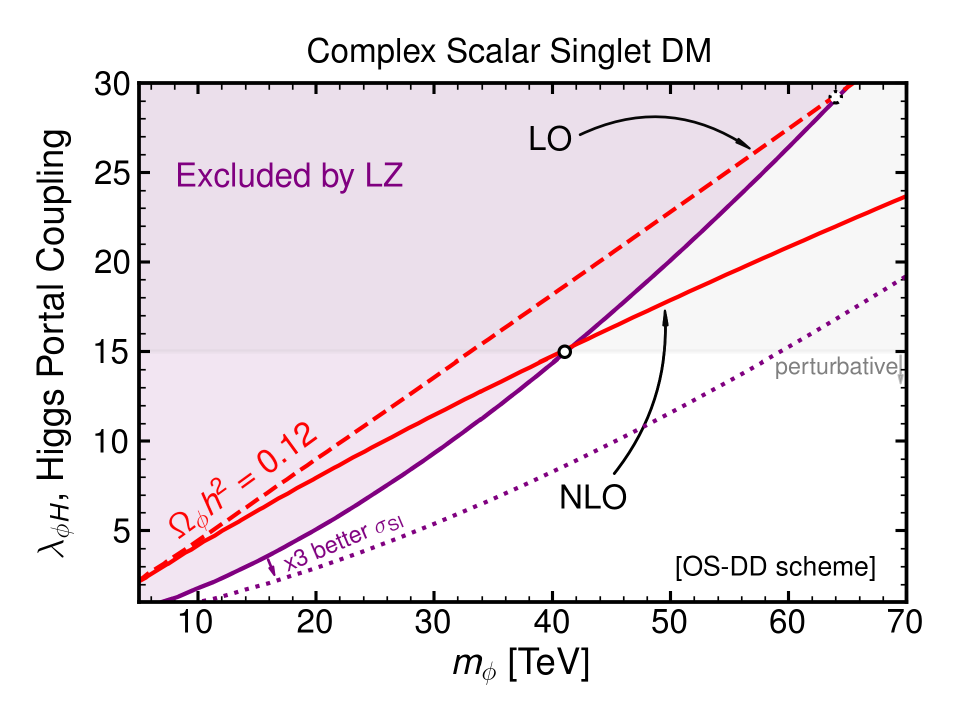}\\
\end{tabular}
\vspace{-0.7cm}
\caption{Parameter space for the singlet scalar dark matter model in the multi-TeV scale region. Left: real scalar, right: complex scalar. Purple corresponds to the recent bounds by LZ at 90\% C.L.~\cite{LZ:2024zvo}. In dashed red we show the values of the coupling that lead to the observed relic abundance at LO in the Higgs portal coupling and in solid red those at NLO. 
Thus, the regions below the solid red curves would lead to dark matter overabundance today. In this figure we show the results in a renormalization scheme where the direct detection cross section is the same at LO and at NLO. A perturbative treatment of the model is expected to be reliable for $\lambda_{\phi H} < 15$, see Eq.~\eqref{eq:25percentlambda}. From the right panel one can see that the entire perturbative parameter space is excluded for the complex scalar case. From the left panel one sees that a three-fold improvement in the sensitivity of direct detection experiments can test the entire perturbative real scalar dark matter TeV-scale window.} 
\label{fig:SingletDM}
\end{center}
\end{figure*}

\begin{figure*}[t]
\begin{center}
\begin{tabular}{cc}
 \includegraphics[width=0.46\textwidth]{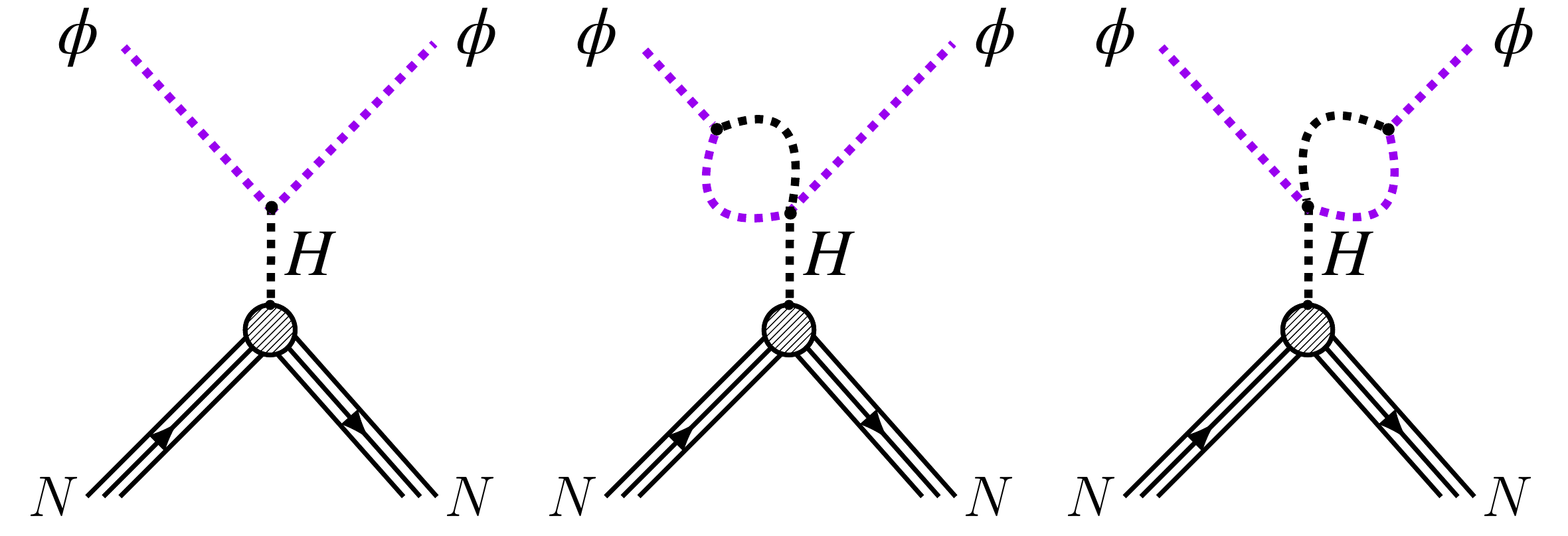}
   \hspace{0.5cm}  &  \hspace{0.5cm} \includegraphics[width=0.45\textwidth]{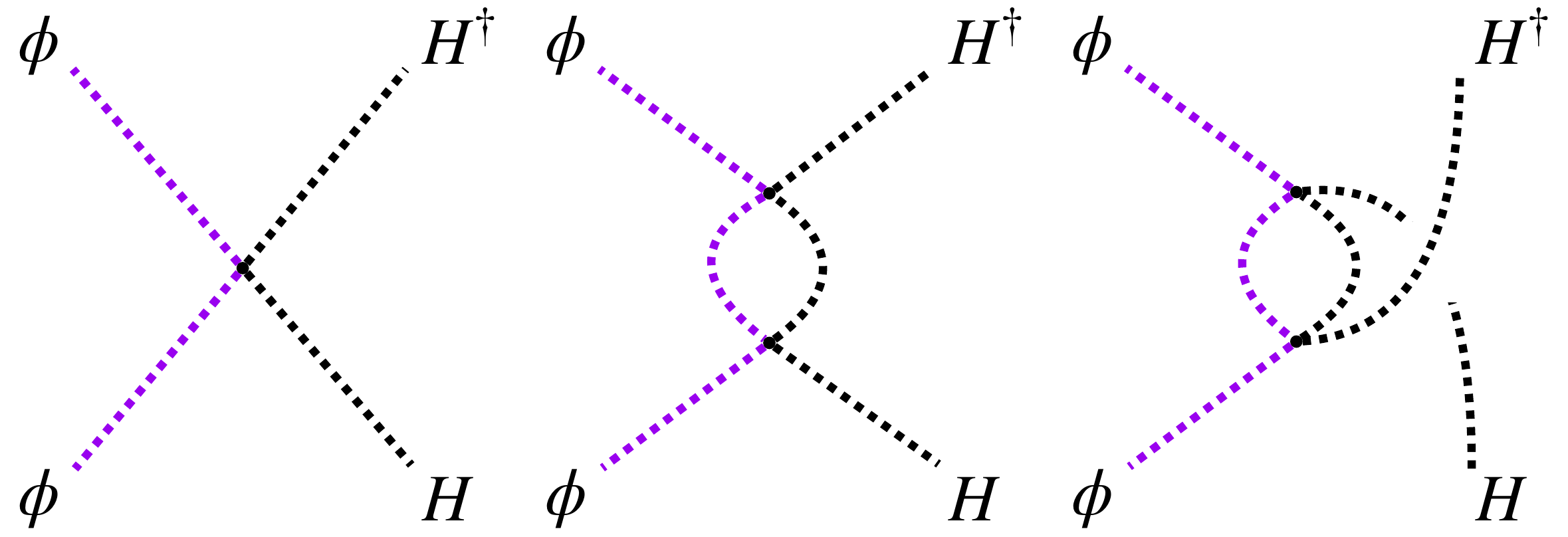}
\end{tabular}
\vspace{-0.2cm}
\caption{LO and NLO diagrams for direct detection (left) and for annihilations (right) in the Higgs portal coupling, $\lambda_{\phi H}$.}
\label{fig:diagrams}
\end{center}
\end{figure*}

\section{Introduction}

\noindent The real and complex scalar singlet dark matter models constitute prototypical scenarios where the relic density stems from the ordinary thermal freeze-out of dark matter (DM) annihilations into Standard Model (SM) particles~\cite{Kolb:1990vq,Gorbunov:2011zz}.  They probably constitute the simplest dark matter scenarios of this type and have been considered in a long series of articles, see e.g.~\cite{Silveira:1985rk,McDonald:1993ex,Burgess:2000yq,Espinosa:2007qk,Andreas:2008xy,Cline:2013gha,Escudero:2016gzx,GAMBIT:2017gge}.
Adding to the Standard Model a massive real or complex scalar singlet, odd under a $Z_2$ symmetry, results in the following extra interaction Lagrangian 
\begin{equation}
{\mathcal L}\,\,\owns \,\,-a\, \lambda_{\phi H} (H^\dagger H) (\phi^\dagger \phi) -\frac{b}{4}\,\lambda_\phi (\phi^\dagger \phi)^2 \,,
\end{equation}
where $a=1/2,1$ and $b=1/6,1$ for the real and complex case, respectively. 

In this model, dark matter production proceeds through the freeze-out of the $\phi\phi$ or $\phi \phi^\dagger$ annihilation into SM particles induced by the Higgs portal interaction $\lambda_{\phi H} $. In addition, dark matter nucleon scattering is mediated by the Higgs boson and is also driven at tree level by the same Higgs portal coupling.

\section{The Singlet Scalar Model at LO}

As well known in the $m_\phi \gg m_h$ regime (which means $\lambda_{\phi H}$ much larger than all SM couplings) the tree level annihilation and direct detection cross-sections are
\begin{align}
    \sigma v_{\rm lab}|_{\rm LO}&= \frac{\lambda_{\phi H}^2}{16\pi m_\phi^2}\,,\label{sigmaannihLO}\\
    \sigma_{\rm SI}|_{\rm LO}&= 
    \frac{\lambda_{\phi H}^2 f_N^2}{4\pi}\frac{m_N^4}{m_h^4}\frac{1}{(m_\phi+m_N)^2}\,,
\end{align}
where here $m_\phi$ is the $\phi$ mass, $m_h$ is the Higgs boson mass, $m_N \simeq 0.938\,{\rm GeV}$ is the nucleon mass, and $f_N\simeq 0.3$ is the Higgs-nucleon coupling. 

Considering that the real $\phi$ particle makes up all of the dark matter in the Universe and for $m_\phi > 1\,{\rm TeV}$, the recent limit from the LZ experiment~\cite{LZ:2024zvo} (see also the results from XENONnT~\cite{XENON:2025vwd} and PandaX~\cite{PandaX:2024qfu}) can be translated into a bound on the Higgs portal coupling of:
\begin{align}
    \lambda_{\phi H} \lesssim 10 \, \left(\frac{m_\phi}{30\,{\rm TeV}}\right)^{3/2}\,.
\end{align}
In addition, thermal freeze-out roughly requires $\left<\sigma v_{\rm lab}\right> \simeq 3\times 10^{-26}\,{\rm cm^3/s}$ which in the TeV scale in turn implies 
\begin{align}
    \lambda_{\phi H} \simeq  10 \, \left(\frac{m_\phi}{30\,{\rm TeV}}\right)\,.\quad 
\end{align}
Figure~\ref{fig:SingletDM} shows these results more precisely. 
From these numbers one can clearly see that the model is pushed to the multi-TeV regime and that the couplings this implies are very large. In this context, it is important to investigate the perturbativity of the model and this is the goal of this work. In what follows, we calculate the NLO corrections to the dark matter nucleon scattering as well as to the annihilation to understand better what is the region of parameter space which is still allowed perturbatively. 

\section{The Singlet Scalar Model at NLO}

In Figure~\ref{fig:diagrams} we show the NLO diagrams that contribute to the dark matter nucleon scattering (left) and to the annihilation (right). We regularize the divergences using dimensional regularization and consider three different renormalization schemes. The comparison of the results in the three schemes will allow to better understand what are the uncertainties that hold at NLO order. These three schemes are: 
\begin{enumerate}
    \item \textbf{MSbar}: The most commonly used scheme and amounts to subtract the $1/\epsilon$ pole in dimensional regularization and the $\gamma_E$ and $4\pi$ constants.  
    \item \textbf{OS-DD}: Our prescription here is to make the direct detection cross section to be the same at both LO and NLO, that is: $\sigma_{\rm SI}|_{\rm NLO}=\sigma_{\rm SI}|_{\rm LO}$. This is a type of ``on-shell" scheme and brings all the NLO corrections to the freeze-out calculation.  
    \item \textbf{OS-FO}: Our prescription here is the opposite one, requiring the annihilation cross section to be the same at LO and at NLO at the kinematical point $s = 4m_\phi^2$, $t = u = -m_\phi^2$. This implies that for the dominant s-wave contribution: $\sigma v_{\rm lab}|_{\rm NLO}=\sigma v_{\rm lab}|_{\rm LO}$. This brings the entire effect of radiative corrections into the direct detection cross section.
\end{enumerate}
We define the quantity $\Delta_{\rm NLO} \equiv {\rm Rate}|_{\rm NLO}/{\rm Rate}|_{\rm LO} - 1$ which will be relevant for freeze-out ($\text{FO}$) and direct detection ($\text{DD}$). Upon explicit calculation we find the following results for the s-wave contribution of the annihilation cross section $\phi \phi^{(\dagger)} \to H H^\dagger $:
\begin{subequations}\label{eq:NLO_FO}
\begin{align}
&\text{NLO corrections to Freeze-Out} \!\!\!\!\!\!\!\!\!\!\!\!\!\!\!\!\!\!\!\!\!\!\!\!\!\!\!\!\!\!\!\!\!\!\!\!\!\!\! &\nonumber \\
&[\overline{\rm MS}\,\mu  = m_\phi] &  \Delta_{\rm NLO}^{\rm FO} &\simeq-0.0155\, \lambda_{\phi H} \,,   \\
&[\text{OS-DD}] & \Delta_{\rm NLO}^{\rm FO} &\simeq +0.035\,(1-x_h)^2\, \lambda_{\phi H}\,, \\
 &[\text{OS-FO}] &  \Delta_{\rm NLO}^{\rm FO} &\equiv 0 \,,
\end{align}
\end{subequations}
where $x_h = m_h/m_\phi$. We note that for the $\overline{\rm MS}$ scheme and for the relevant $m_\phi \gg m_h$ case there is only a very mild $m_\phi$ sensitivity on the results. By definition, the NLO correction for the OS-FO scheme vanishes. For $\phi$-nucleon interactions as relevant for direct detection we find:
\begin{subequations}\label{eq:NLO_DD}
\begin{align}
&\text{NLO corrections to Direct Detection} \!\!\!\!\!\!\!\!\!\!\!\!\!\!\!\!\!\!\!\!\!\!\!\!\!\!\!\!\!\!\!\!\!\!\!\!\!\!\!\!\!\!\!\!\!\!\!\! \!\!\!\!\!\! \!\!\!\!\!\! &\nonumber \\
&[\overline{\rm MS}\,\mu  = m_\phi] &  \Delta_{\rm NLO}^{\rm DD} &\simeq -0.051 \,(1-x_h\pi/2) \,\lambda_{\phi H}\,,   \\
&[\text{OS-DD}] & \Delta_{\rm NLO}^{\rm DD} &\equiv 0\,, \\
 &[\text{OS-FO}] &  \Delta_{\rm NLO}^{\rm DD} &\simeq -0.035\,(1-x_h)^2\, \lambda_{\phi H}\,,
\end{align}
\end{subequations}
where we have worked to leading order in $x_h = m_h/m_\phi$ (see Appendix~\ref{app:NLOcalc} for the exact formulae). Note that $\Delta_{\rm NLO}^{\rm FO}-\Delta_{\rm NLO}^{\rm DD}$ is scheme independent and this reduces the scheme dependence of the lower bound on $m_\phi$.

\textit{Relic abundance calculation -- }
To obtain the relic abundance we consider the usual procedure~\cite{Gondolo:1990dk} and follow explicitly Ref.~\cite{Steigman:2012nb} (which we used also for the tree level results). We take the degrees of freedom in the plasma and their relevant derivatives from~\cite{Laine:2015kra}. We solve the Boltzmann equation numerically taking into account both s-wave and p-wave annihilations and we use as an initial condition for the $\phi$ abundance the thermal number density. We do not take into account thermal effects~\cite{Biondini:2025ihi} or Sommerfeld enhancement~\cite{Garcia-Cely:2015khw} which are both numerically negligible. We solve the system from $T = m_\phi/15$ down to $T = 10^{-5}\,m_\phi $ using a stiff differential equation solver in \texttt{Mathematica} with a high precision setting. We contrast the final abundance with the measured value as reported by Planck $\Omega_{\rm DM} h^2 = 0.120\pm 0.001$~\cite{Planck:2018vyg}. 

\textit{Direct detection bounds -- }
To translate the limits from the LZ experiment into the $\lambda_{\phi H}$ coupling we assume the standard halo model as in~\cite{LZ:2024zvo} and use $f_N = 0.308 \pm 0.018 $ from~\cite{Hoferichter:2017olk}. We note that recent lattice results~\cite{Gupta:2021ahb} seem to indicate that the coupling may be slightly larger and in agreement with the findings of~\cite{Alarcon:2011zs}. In practice, see~\cite{Cousins:1991qz}, we derive the limit on $\lambda_{\phi H}$ by simply using the 90\% C.L. result on $\sigma_{\rm SI}$ from~\cite{LZ:2024zvo} and using the mean value of $f_N = 0.308$. By bootstrapping the LZ results, doing a profile likelihood on $\lambda_{\phi H}$ and varying simultaneously $f_N$, we explicitly checked that the (small) uncertainty on $f_N$ has a negligible impact on the 90\% C.L. direct detection limit on $\lambda_{\phi H}$.

\begin{table}[t]
\renewcommand{\arraystretch}{1.2}
\setlength{\arrayrulewidth}{.25mm}
\centering
\small
\setlength{\tabcolsep}{0.18 em}
\begin{tabular}{ c | c | c | c | c  | c }
\hline\hline
\multirow{2}{*}{\textbf{Scheme}$\,$}        	&	\multirow{2}{*}{\textbf{$m_\phi$}$/\lambda_{\phi H}$}	&  \multicolumn{2}{c|}{$\,\,\,\,\,\,\,\,\,$\textbf{Real Scalar}$\,\,\,\,\,\,\,\,$} &  \multicolumn{2}{c}{$\,\,$\textbf{Complex Scalar} $\,\,$}	  \\  \cline{3-6}
    			& &  $\,\,\,\,\,\,$ DD $\,\,\,\,\,\,$  &  Pert. 				&  $\,\,\,\,\,\,$ DD $\,\,\,\,\,\,$ &  Pert.       \\ \hline
\multirow{2}{*}{\textbf{LO}$\,$}       & $m_\phi$ &  \,\, 31 \,\,  &  47 				&  \,\, 64 \,\,  &  33 		  \\ \cline{2-6}
                                        & $\lambda_{\phi H}$ &  \,\, 9.9 \,\,  &  15 				&  \,\, 29 \,\,  &  15 		  \\ \hline
\multirow{2}{*}{\textbf{MSbar}$\,$}   & $m_\phi$&  \,\, 22 \,\,  &  42 				&  \,\, 26 \,\,  &  29 		  \\ \cline{2-6}
                                       & $\lambda_{\phi H}$ &  \,\, 7.2 \,\,  &  15 				&  \,\, 13 \,\,  &  15 		  \\ \hline
\multirow{2}{*}{\textbf{OS-DD}$\,$}   & $m_\phi$&  \,\, 24 \,\,  &  58 				&  \,\, 41 \,\,  &  41 		  \\ \cline{2-6}
                                      & $\lambda_{\phi H}$ &  \,\, 6.9 \,\,  &  15 				&  \,\, 15 \,\,  &  15 		  \\ \hline
\multirow{2}{*}{\textbf{OS-FO}$\,$}   & $m_\phi$&  \,\, 23 \,\,  &  47 				&  \,\, 32 \,\,  &  33 		  \\ \cline{2-6}
                                        & $\lambda_{\phi H}$ &  \,\, 7.1 \,\,  &  15				&  \,\, 14 \,\,  &  15 		  \\ \hline\hline
\end{tabular}
\caption{Summary of the results at LO and at NLO for the three different renormalization schemes as described in the text. DD refers to the bounds from direct detection at 90\% C.L. and should be taken as lower bounds for $m_\phi$. Pert. refers to the values of the maximum mass that can yield the observed dark matter abundance for the value of the coupling $\lambda_{\phi H} \simeq 15$ up to which we expect perturbativity to still hold, see Eq.~\eqref{eq:25percentlambda}.}
\label{tab:bounds}
\end{table}

\vspace{-0.5cm}
\section{Results}
\vspace{-0.3cm}

\noindent \textbf{Lower bounds on $m_\phi$ at NLO:} Our main results are visually summarized in Figure~\ref{fig:SingletDM}. There we consider the OS-DD scheme which has the nice feature that it leads to the same direct detection constraint at LO and NLO. From the right panel one observes that the bound for the complex case goes from $m_\phi \simeq 64\,{\rm TeV}$ at LO to $m_\phi \simeq 40\,{\rm TeV}$ at NLO, while from the left panel one sees that for the real case it goes from $m_\phi \simeq 31\,{\rm TeV}$ (LO) to $m_\phi \simeq 24\,{\rm TeV}$ (NLO).

\textbf{Scheme dependence of the lower bound:} Figure~\ref{fig:SingletDM_3cases} in the Appendix contains the results for the three NLO schemes. Numerically, the results are summarized in Table~\ref{tab:bounds}. By looking at the numbers one can see that the variation of the lower bound for the real case is mild, within the $22-24\,{\rm TeV}$ range. This is good because it is a sign that the model is still well within the perturbative regime. Thus, a value below $\sim 20 \, {\rm TeV}$ appears to be firmly excluded in the real case. Instead, for the complex scalar, the lower bound largely varies from 26 TeV to 40 TeV, a sign that in this case perturbativity is already under question.


\textbf{Perturbativity:} The key question is up to which value of the mass do we stand still in the perturbative regime? This question holds for both the real and complex cases. It is known that to look at the convergence of the perturbative series by comparing the relative size of the various terms for a given observable in a single scheme can lead to misleading results, especially if we stop at the NLO, so that one can only compare the NLO contribution to the LO one. For instance, a calculation of the $hh\rightarrow h h$ scattering up to NNLO reveals that the ratio between the NNLO to NLO can be larger than the NLO to LO ratio and that the convergence of the perturbative series can be much better for appropriate choices of the scheme, see e.g.~\cite{Riesselmann:1996is}.\footnote{For an ``unappropriate" choice of the scheme one will easily find that the series does not converge already for quite low masses (i.e.~couplings), a result that can be easily misleading. Instead for a ``too appropriate" choice of the scheme, one will instead find an apparent breakdown of perturbativity for much larger masses, which cannot be trusted either because for such masses for instance the $\beta$ function series of the quartic coupling would be clearly not converging.} From this discussion, and comparison with lattice results, in $\lambda \phi^4$ theories it appears that a realistic perturbativity condition above which we should not go is to be found at the level of the $\beta$ function, requiring that the next to leading log contribution to it is not larger than 25\% of the leading log one, see e.g.~\cite{Riesselmann:1996is,Luscher:1988gc,Hambye:1996wb}. Although perturbativity of course somewhat depends on the interaction and observable considered, this is the reasonable prescription we will use. On the other hand for 50\% instead one expects to be already deep in the non-perturbative regime. 
To address the perturbativity question we will therefore evaluate the $\beta$ function of the portal coupling up to next-to-leading log order. To this end we use the Mathematica program \texttt{SARAH}~\cite{Staub:2008uz,Staub:2015kfa} to automatically compute the one-loop and two-loop beta functions. Considering only the $\lambda_{\phi H}$ coupling, the results are as follows for the real case,
\begin{align}
    \frac{d\lambda_{\phi H}}{d\log \mu} = \frac{1}{16\pi^2} \left[4\lambda_{\phi H}^2\right] + \frac{1}{(16\pi^2)^2}\left[-\frac{21}{2}\lambda_{\phi H}^3\right]
\,,    \label{betafunction}
\end{align}
whereas in the complex case one gets the same up to a small difference: the factor $-21/2$ is replaced by $-11$ (in agreement with~\cite{Brod:2020lhd,Bandyopadhyay:2025ilx}). 

\noindent Applying the 25\% prescription above gives then
\begin{align}
    \lambda_{\phi H}(\mu) < \frac{32\pi^2}{21}\simeq 15\, \quad \text{[Perturbativity]}\,.
    \label{eq:25percentlambda}
\end{align}
Plugging in this value into the relic density constraint in the three schemes leads to the upper values of the masses quoted in Table~\ref{tab:bounds} for both the real and complex cases. For the complex case, since the direct detection lower bound on the mass is above or very close to these values, one concludes that the complex case is ruled out in its perturbative regime. For the real case as it is well below one concludes that there is still an allowed perturbative regime range that goes from the 20 TeV mass scale quoted above up to possibly about 40-50~TeV.\footnote{The fact that in the OS-FO and $\overline{\hbox{MS}}$ schemes the direct detection constraint turns back at already $\sim 34$~TeV and $\sim 26$~TeV respectively, see Fig.~\ref{fig:SingletDM_3cases} (as a result of the fact that the NLO correction is large and has opposite sign with respect to the LO one) does not necessarily mean that we are already lying in the non-perturbative regime, see footnote 1 above, but clearly it is the sign that non-perturbativity is close.} Given the dashed purple lines in the figures we can also conclude that if the upper bound on the direct detection cross section improves by a factor of three the full perturbative regime would be ruled out also for the real scalar singlet dark matter case.

 \textbf{Unitarity bounds:} Partial wave unitarity constraints have been considered specifically for the singlet scalar dark matter model, see~\cite{Goodsell:2018tti} and also~\cite{Griest:1989wd,Goodsell:2020rfu}. Considering the regime where $\lambda_{\phi H} \gg \lambda_\phi,\lambda_H$ the dominant high energy scattering is $\phi H \to \phi H$ which in order to be unitary at tree level requires $|\lambda_{\phi H}| < 8\pi \simeq 25$. This value is higher than the realistic perturbativity bound obtained above in Eq.~\eqref{eq:25percentlambda}, which could have been anticipated because the unitarity bound is to be expected to arise already in the deeply non-perturbative regime.

\textbf{Vacuum stability and the sign of the Higgs portal coupling:} The range of masses which appears to be still allowed in a perturbative regime for the real case holds only for a positive value of the Higgs portal coupling, $\lambda_{\phi H}>0$. A negative coupling is excluded in the perturbative regime because the minimum value of the portal coupling that direct detection constraints implies, $|\lambda_{\phi H}|\gtrsim 7$, is too large not to destabilize the vacuum for perturbative values of the $\lambda_\phi$ quartic coupling. Vacuum stability requires $\lambda_{\phi H}> -\sqrt{2\lambda_H\lambda_\phi/3}$ (see e.g.~\cite{Falkowski:2015iwa}), which, given that the SM quartic coupling $\lambda_H$ is equal to $0.13$ would require $\lambda_\phi \gg 100$.

\textbf{Possible effect of a large $\phi$ quartic coupling:} A potentially large $\lambda_\phi$ quartic coupling could change the annihilation and direct detection rates and possibly alter the $m_\phi$ lower bound. However, by explicitly calculating the NLO corrections from it we find that the $\lambda_\phi$ correction has the opposite sign to that of $\lambda_{\phi H}$. For example, for $m_\phi = 20\,{\rm TeV}$ one finds $\Delta_{\rm NLO}^{\rm FO}-\Delta_{\rm NLO}^{\rm DD}=+0.035\, \lambda_{\phi H}-0.012 \,\lambda_{\phi}$. Since they have opposite signs and both couplings must be positive the lower bound on $m_\phi$ cannot be reduced by radiative corrections induced by $\lambda_\phi$.

\textbf{Appearance of Landau poles:} 
The large values of the Higgs portal coupling that direct detection constraints imply, and the fact that only positive values of the Higgs portal are still allowed, unavoidably raise the question of possible Landau poles at a scale close to $m_\phi$. As for large SM quartic coupling, Landau poles show up at the one loop RGE level because the beta function is positive. As for the SM quartic coupling too, at the two loops instead, a fixed point arises due to the negative two-loop contribution to the $\beta$ function. To estimate what is the scale where a Landau pole is expected to show up we will run the coupling from the two-loop $\beta$ function given in Eq.~(\ref{betafunction}) and, as in \cite{Hambye:1996wb} for the SM case, adopt the prescription that it arises when the portal coupling reaches half the value of the portal coupling fixed point value (motivated by lattice results \cite{Agodi:1994qv,Cea:1996af,Luscher:1988gc}). Such condition is also close to imposing that the two-loop contribution to the $\beta$ function is about half of the one-loop one. The left panel of Figure~\ref{fig:SingletDM} shows the values of $\lambda_{\phi H}$ at the $m_\phi$ scale that leads to a Landau pole at a scale 20 or 100 times larger than $m_\phi$. An upper limit on the Landau pole is obtained if we consider the lower bound on the masses given in Table~\ref{tab:bounds}. In the MSbar/OS-FO/OS-DD schemes this corresponds to 5300, 6000, 7400 TeV, respectively. As for the upper end of the perturbative range ($\lambda_{\phi H}=15$) we get a Landau pole arising only a factor $\simeq 8$ larger than the $\phi $ mass corresponding to $340,\,380,\,460$ TeV in the three schemes. Consequently, the framework requires that new physics must manifest itself below these low scales to calm down the running of $\lambda_{\phi H}$.

\section{The $\large{m_\phi \simeq m_h/2}$ resonant region}
Before concluding, we would like to comment on the narrow low mass window  where these dark matter models are viable: the $m_\phi \simeq m_h/2$ region. In this region the annihilation is resonantly enhanced, so that to account for the observed relic density requires much smaller couplings, $\lambda_{\phi H} \sim 10^{-3}$. In~\cite{Binder:2017rgn} it has been  shown that, given such small couplings, the assumption usually made in this region that dark matter particles are in kinetic equilibrium at freeze-out may not be fully legitimate. Using the \texttt{DRAKE} code from the same authors~\cite{Binder:2021bmg} we will calculate the relic abundance with and without imposing kinetic equilibrium. To this end, we consider their simplified approach that amounts to solving a set of two coupled differential equations for the number density and the second moment of the $\phi$ distribution function. Applying the recent LZ direct detection constraints at 90\% C.L. (and also updating the small change of the Higgs boson mass in \texttt{DRAKE}, $m_h =125.2(1)\,{\rm GeV}$) we get the following allowed regions for the real case:
\begin{subequations}
\begin{align}
m_\phi &= (59.2-62.6)\, {\rm GeV}    \quad [\text{Real}/ \text{kinEQ}]\\
m_\phi &= (59.4-62.6)\, {\rm GeV}    \quad [\text{Real}/ \text{nokinEQ-A}]\\
m_\phi &= (60.2-62.6)\, {\rm GeV}    \quad [\text{Real}/ \text{nokinEQ-B}]
\end{align}
\end{subequations}
where the two non-kinetic equilibrium cases bracket the uncertainty on the number density of quarks in the plasma, see~\cite{Binder:2017rgn,Binder:2021bmg}.
Similarly for the complex case we get:
\begin{subequations}
\begin{align}
m_\phi &= (60.2-62.6)\, {\rm GeV}    \quad  \!\![\text{Complex}/ \text{kinEQ}]\\
m_\phi &= (60.3-62.6)\, {\rm GeV}     \!\!\quad[\text{Complex}/ \text{nokinEQ-A}]\\
m_\phi &= (60.7-62.6)\, {\rm GeV}    \quad  \!\![\text{Complex}/\text{nokinEQ-B}]
\end{align}
\end{subequations}
 We note that the upper limit from direct detection experiments for dark matter masses around 60 GeV should improve by a factor of $\sim 12$ in order to completely test this resonance window for the complex case. For the real case it would need to be a factor of $\sim 25$ better and lie near the neutrino floor~\cite{OHare:2021utq}.

\section{Conclusions}
The real and complex scalar singlet dark matter models are perhaps the simplest examples of thermal dark matter candidates. In this work, by calculating the NLO corrections to thermal freeze-out and direct detection we have shown that the recent limits from the LZ experiment push the lower limit for the complex scalar case to a regime where perturbative calculations are no longer reliable (besides restricting further the allowed mass range around the Higgs boson resonance). For the real case this combination brings the lower bound to 20~TeV. Both the small scheme dependence of this lower bound, and the behavior of the $\beta$ function for this value, shows that, there, the theory is still perturbative. For heavier masses  perturbativity quickly does not hold anymore but it is conceivable that the real scalar singlet dark matter model is still perturbative up to 40-50 TeV. An improvement of the direct detection sensitivity by no more than a factor three would fully test this region as well.
Within this region only positive values of the Higgs portal coupling are allowed, which unavoidably leads to Landau poles. The direct detection lower bound on $m_\phi$ implies that such a Landau pole lies below 7400~TeV. 

These conclusions are relevant for any model along which thermal freeze-out occurs through the Higgs portal interaction (more precisely, for any dark matter scenario for which the relic density stems from the freeze-out of an annihilation induced by a $H^\dagger H \,{\rm DM \overline{DM}}$ interaction). This includes scenarios such as scalar electroweak multiplets\footnote{For electroweak multiplets the problem is avoided if the annihilation goes dominantly into SM gauge bosons \cite{Cirelli:2005uq,Cirelli:2009uv,Bottaro:2021snn,Bloch:2024suj}. These conclusions also do not hold for instance for the inert scalar doublet model (see e.g.~\cite{Barbieri:2006dq,LopezHonorez:2006gr,Hambye:2009pw}) if in particular one tunes the various quartic couplings together so that the annihilation goes dominantly to the SM charged scalar components $h^+ h^-$, with smaller rate into $(h^2+2hv)$, see e.g.~\cite{Hambye:2009pw}.} and asymmetric dark matter models where the Higgs portal ensures efficient annihilations so that an asymmetry dictates the dark matter relic density, see e.g.~\cite{Ibe:2011hq,Dhen:2015wra}.

\acknowledgments
We would like to thank Chandan Hati for discussions. 
The work of TH is supported by the Belgian IISN convention 4.4503.15 and by the Brussels Laboratory of the Universe - BLU-ULB.

\appendix
\section{NLO results for the three Schemes}\label{app:NLOcalc}

\begin{figure*}[t]
\begin{center}
\begin{tabular}{cc}
 \includegraphics[width=0.49\textwidth]{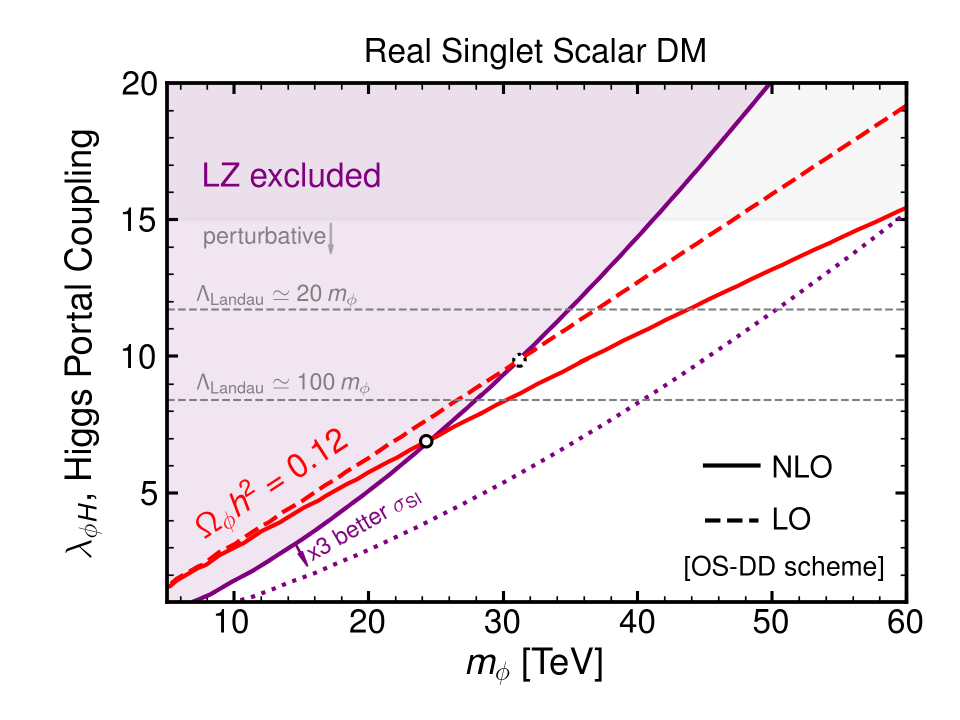}    &   \includegraphics[width=0.49\textwidth]{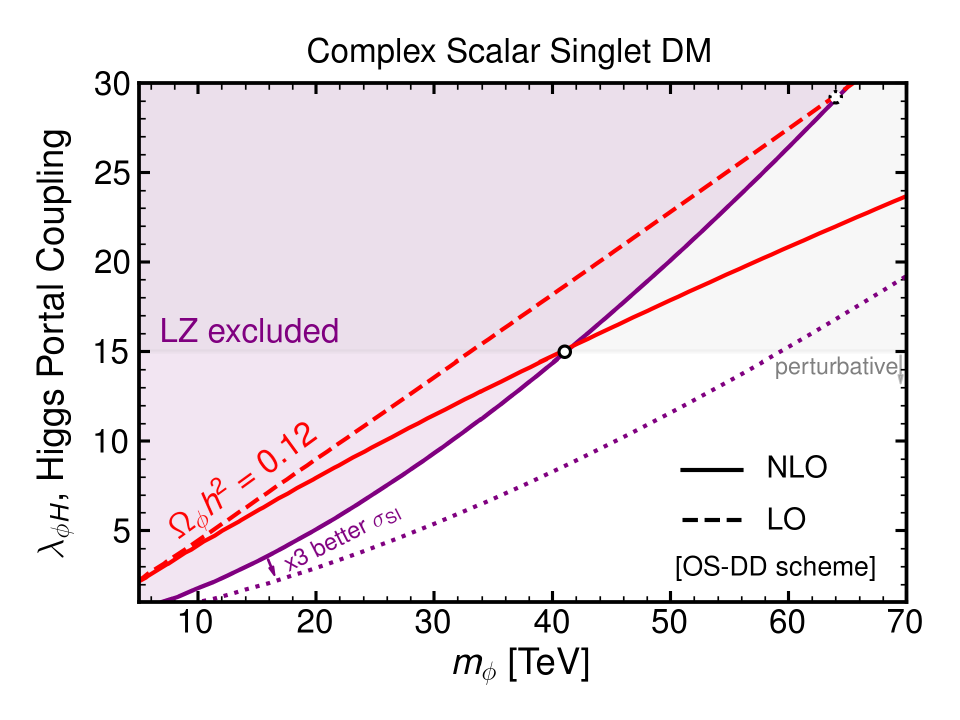}\\
 \includegraphics[width=0.49\textwidth]{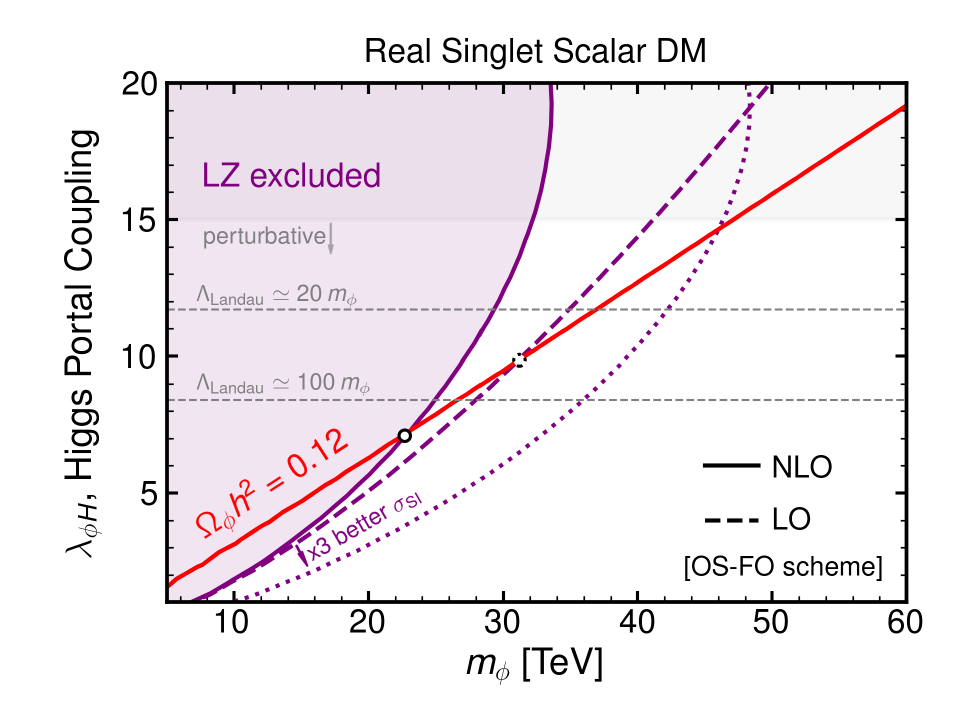}   &   \includegraphics[width=0.49\textwidth]{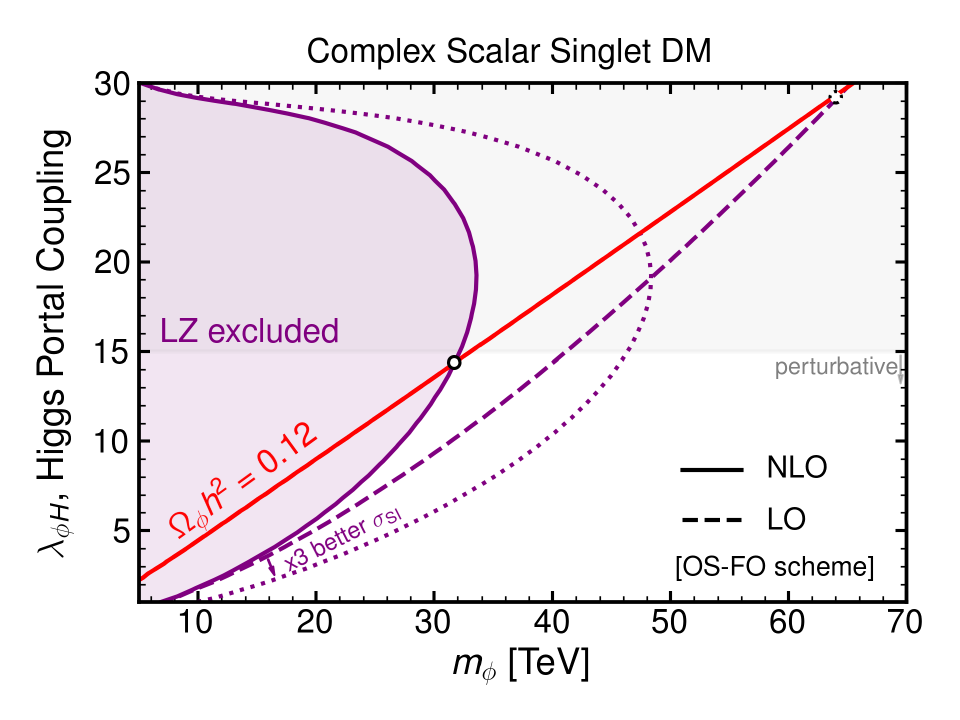} \\
  \includegraphics[width=0.49\textwidth]{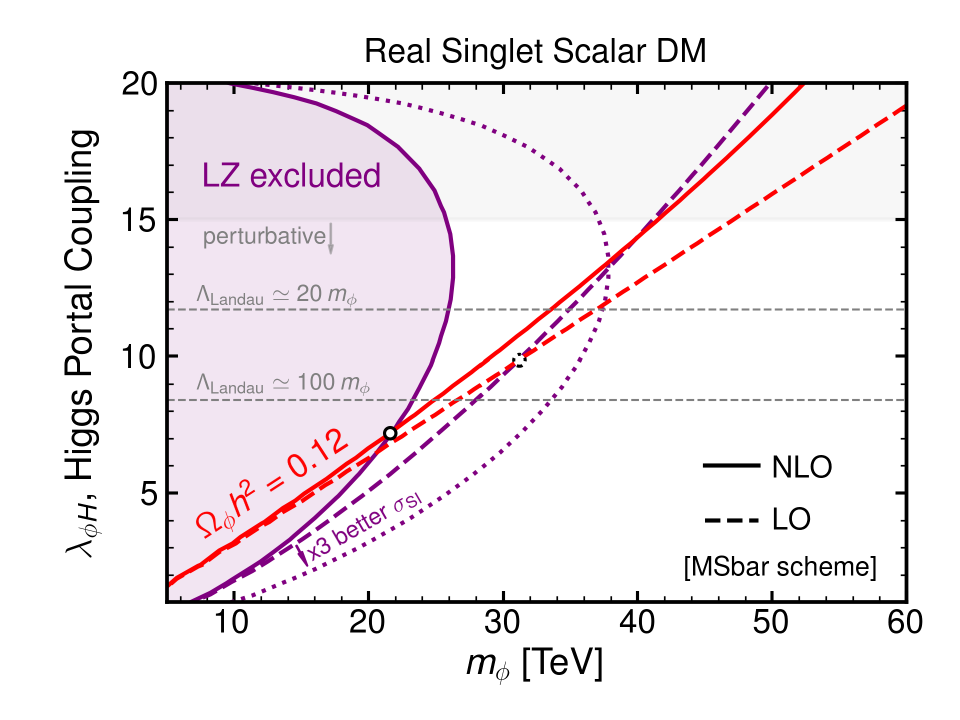}   &   \includegraphics[width=0.49\textwidth]{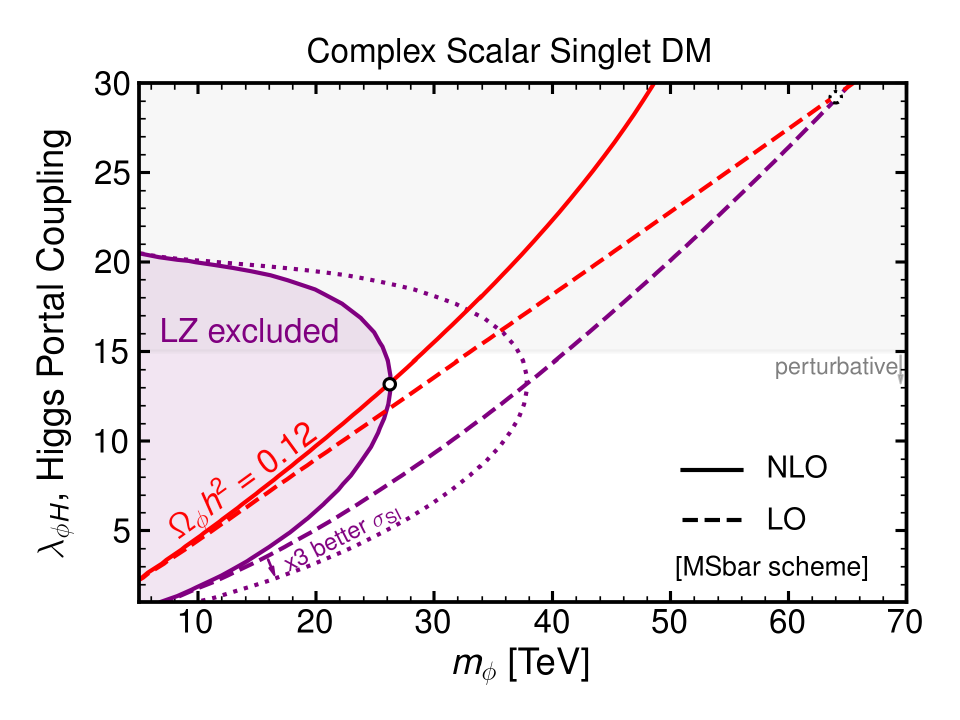} \\
\end{tabular}
\vspace{-0.7cm}
\caption{Same as Figure~\ref{fig:SingletDM} but for the three renormalization schemes employed. Top (OS-DD), middle (OS-FO), bottom (MSbar).}
\label{fig:SingletDM_3cases}
\end{center}
\end{figure*}

For $\phi \phi^{(\dagger)} \to H H^\dagger$ annihilations the relative NLO rate can be written as:
\begin{equation}
    \Delta^{\rm FO}_{\rm NLO} =\frac{\lambda_{\phi H}}{8 \pi^2}\int_{t_0}^{t_1} A \,dt\,,
\end{equation}
where $A$ is the ratio of the amplitudes squared, and $t_{0,1}$ are the usual $t$ Mandelstam variable boundaries for a 2 to 2 cross section. Defining $\lambda(m_\phi^2,m_h^2,s)\equiv (m_\phi^2+m_h^2-p^2)^2-4m_\phi^2m_h^2 $, in the three schemes one gets
\begin{subequations}
\begin{align}
    A_{\overline{\rm MS}}&=f_1(m_\phi^2, m_h^2,t) +f_1(m_\phi^2, m_h^2,u) \,, \\
    A_{\rm OS-FO}&= 0 \,,\\
    A_{\rm OS-DD}&=f_1(m_\phi^2, m_h^2,t) +f_1(m_\phi^2, m_h^2,u)\nonumber\\
    &\,\,\,\,\,\,\,\,-2 f_{2}(m_\phi^2,m_h^2,m_\phi^2)\,.
\end{align}
\end{subequations}

For the direct detection cross section we have:
\begin{align}
    \Delta^{\rm DD}_{\rm NLO} = \frac{\lambda_{\phi H}}{4\pi^2} B  \,,
\end{align}
with $B$ being:
\begin{subequations}
\begin{align}
 \!\!\! B_{\overline{\rm MS}}&=f_{2}(m_\phi^2, m_h^2,m_\phi^2) \,, \\
 \!\!\!    B_{\rm OS-FO}&= f_{2}(m_\phi^2, m_h^2,m_\phi^2)-f_1(m_\phi^2,m_h^2,-m_\phi^2)\,,\\
 \!\!\!    B_{\rm OS-DD}&=0\,.
\end{align}
\end{subequations}
The two functions are
\begin{align}
&f_{1,2}(m_\phi^2,m_h^2,p^2)=-2+\frac{1}{2}\log\Big(\frac{m_\phi^2}{\mu^2}\Big)+\frac{1}{2}\log\Big(\frac{m_h^2}{\mu^2}\Big)\nonumber\\
&+\frac{m_\phi^2-m_h^2}{2p^2}\log\Big(\frac{m_\phi^2}{m_h^2}\Big)+g_{1,2}(m_\phi^2,m_h^2,p^2)\,,
\end{align}
with
\begin{align}
&g_{1}(m_\phi^2,m_h^2,p^2)=\frac{\sqrt{\lambda(m_\phi^2,m_h^2, p^2)}}{2 p^2} \\
&\times \log\bigg({\left[m_\phi^2+m_h^2-p^2-\sqrt{\lambda(m_\phi^2,m_h^2, p^2)}\,\right]^2}\bigg/[{4m_\phi^2 m_h^2}]\bigg)\nonumber \,,
\end{align}
and
\begin{align}
&g_{2}(m_\phi^2,m_h^2,p^2)=\frac{\sqrt{-\lambda(m_\phi^2,m_h^2, p^2)}}{p^2}\\
&\times  \hbox{Arctan}\bigg({\sqrt{-\lambda(m_\phi^2,m_h^2, p^2)}}\bigg/{[{m_\phi^2 +m_h^2-p^2}]}\bigg)\nonumber\,.
\end{align}

\bibliography{biblio}

\end{document}